\documentclass{mn2e}
\usepackage{graphics}
\usepackage{epsfig}
\usepackage{lscape}

\def\beq{\begin{equation}}                          
\def\eeq{\end{equation}}                              
\def\beqa{\begin{eqnarray}}                         
\def\eeqa{\end{eqnarray}}                             
\def\beqan{\begin{eqnarray*}}                      
\def\eeqan{\end{eqnarray*}}                          

\newbox\grsign \setbox\grsign=\hbox{$>$} \newdimen\grdimen \grdimen=\ht\grsign
\newbox\simlessbox \newbox\simgreatbox
\setbox\simgreatbox=\hbox{\raise.5ex\hbox{$>$}\llap
     {\lower.5ex\hbox{$\sim$}}}\ht1=\grdimen\dp1=0pt
\setbox\simlessbox=\hbox{\raise.5ex\hbox{$<$}\llap
     {\lower.5ex\hbox{$\sim$}}}\ht2=\grdimen\dp2=0pt

\title[Multi-Phase Dusty Gas in the Center of NGC 4278]
   {Multi-Phase Dusty Gas in the Center of NGC 4278}

\author[Y. -P. Tang, Q.-S. Gu, S.-N. Zhang, B.-T. Tang]
       {Yuping Tang$^{1}$\thanks{E-mail: tangyping@gmail.com},
        Qiusheng Gu\thanks{E-mail: qsgu@nju.edu.cn},
       Shuinai Zhang,
       Baitian Tang
\\
   $^{1}$Department of Astronomy, Nanjing University, Nanjing 210093, P. R. China\\
   Key Laboratory of Modern Astronomy and Astrophysics
(Nanjing University), Ministry of Education, Nanjing 210093,
China\\}

\begin{document}

\maketitle

\begin{abstract}

We present the Spitzer spectroscopic mapping observations toward the
central kpc region of the nearby elliptical LINER galaxy, NGC 4278,
by using the Infrared Spectrograph (IRS). These observations reveal
rich mid-IR emission features of extended ionized gas, warm
molecular hydrogen and dust. Different phases of gas and dust are
closely related and belong to a same elongated feature. We further
study properties of multi-phase dusty gas to uncover the underlying
mechanism of ionization and excitation. The band ratio and intensity
of PAH features in the central region might reflect modified size
distribution resulted from selective destruction. H$_{2}$ S(0)-S(7)
pure rotational lines of molecular hydrogen show excessive intensity
and moderately high excitation temperature comparing with photon
dissociation region (PDR). A strong and extended [SiII] emission
line is detected, which could be a sign of reduced depletion of
silicon in interstellar dust. We also discover an extended high
ionization region associated with enhanced H$_{2}$ S(1) emission. We
conclude that a shock-heating component is required to account for
observed emission characteristics, which could be triggered by
cloud-cloud interactions during accretion of cold gas from the large
HI disk.

\end{abstract}

\begin{keywords}
galaxies: active - galaxies: elliptical and lenticular,cD -
galaxies: nuclei - infrared: galaxies - NGC 4278
\end{keywords}

\section{Introduction}

A uniform, old stellar population plus extremely hot gas ($T \sim
10^6 - 10^7 K$) were considered the only components of elliptical
galaxies historically. However, recent studies confirm that cold and
warm diffuse matter are not rare
--- although generally small in amount --- in these systems. The
SAURON integral-field Survey of Early-Type Galaxies (Sazi et al.
2006) reported a 66 percent detection of extended emission gas for
ellipticals, with a wide variety of spacial distributions and
kinematic behaviors. Neutral hydrogen observations show that plenty
of extended neutral hydrogen gas is clearly present in elliptical
galaxies (Morganti et al. 2006; Noordermeer 2006). Moreover, even
cooler interstellar medium (ISM) including dust and molecular gas
has been detected at a modest level(Knapp et al. 1989; Lauer et al.
2005; Wiklind et al.1995; Sage et al. 2007).

The origin of both warm and cold matter in nearby elliptical
galaxies still remains controversial, clarifying this question is
important for us to understand the ongoing physical processes that
drive the evolution of this class of systems. In many cases, the
warm and cold gas are linked with each other in nearby ellipticals
(Morganti et al. 2006; Serra et al. 2008), suggesting the two
components might be different phases of a same structure. For
emission gas, line ratios typical of LINERs are usually found in
nearby ellipticals (Ho et al. 1997; Filippenko, 2003). A variety of
mechanisms have been applied to interpret its energy source. Besides
photoionization by active nucleus and low level residual star
formation, possibilities of photoionization by an old, hot stellar
population(Binette 1994; Macchetto et al. 1996), cooling flow or
thermal interaction with hot gas (Sparks et al. 1989; de Jong et al.
1990; Fabian, 1994) and heating via shocks (Doptia \& Sutherland,
1995) were also discussed and examined by previous authors, while a
conclusive answer has not yet been reached. On the other hand, the
lack of any significant correlation between the cold diffuse matter
and stars on both mass and kinematics in nearby ellipticals might be
a sign of external origin for the former content(Knapp et al. 1985;
Goudfrooij \& de Jong, 1994; Oosterloo et a. 2002), although the
continuous thermal and dynamical evolution of ellipticals, such as
heating of stellar ejecta through random collision, might assist to
obscure such a relationship.

The nearby ($D=16.1 Mpc$; Tonry et al. 2001), isolated elliptical
galaxy NGC 4278 is a rare sample of ellipticals with detections of
multi-phases ISM, thus offering a significant opportunity to study
the evolution of gaseous matter in such environment. As among the
first ellipticals detected of cold gas, NGC 4278 has long been known
for its massive ($\sim 10^8 M_{\odot}$), regular HI disk extending
beyond $5 kpc$. Further more, the detection of CO emission is
recently reported by Combes et al. (2007). Strong emissions produced
by ionized gas are revealed by the SAURON survey (Sazi et al. 2006),
which kinematically coincides well with the HI disk (Morganti et al.
2006) and show asymmetrical sub structures with high ionization.  In
optical images, NGC 4278 shows large scale dust patches located
north and north-west of the galaxy center (Lauer et al. 2005), which
might contribute to excess far-infrared emission detected by IRAS
(Knapp et al. 1989) and ISO (Temi et al. 2004) infrared space
telescopes. Besides, mid-infrared extended, non-stellar emission
possibly related to reprocessing of UV/optical radiation by
polycyclic aromatic hydrocarbons (PAHs) is shown in Spitzer IRAC
image after removing the stellar contribution (Tang et al. 2009).

In this work, we present mid-infrared spectral mapping observations
of Spitzer Space Telescope toward the central kpc region of NGC 4278
. There are several reasons why spatially resolved mid-infrared
spectroscopy in such area is interesting. First, it allows us to
carry out analysis on spectra without suffering from serious dust
extinction; second, the uniform spectral and spacial stellar
background of elliptical galaxy provide us chances to identify
activities characterized by low surface brightness; third,
mid-infrared spectra contain emission features from multi-phase ISM,
including dust, warm molecular hydrogen as well as ionized gas, thus
enable us to explore the underlying heating mechanism from various
aspects. Our ultimate goal is to understand the origin and evolution
of cold and warm matter in this LINER galaxy.

This paper is organized as follows: The observation and data
reduction are described in Section 2. The emission line maps are
shown in Section 3. We explore the physical properties of cold and
warm matter from observations in Section 4 and carry out discussion
on possible energy source behind the diffuse matter in Section 5. In
Section 6 we draw our conclusions.


\section{Observation and Data Reduction}

  We download the IRS Basic Calibrated Data (BCD) of NGC 4278 from the archive of Spitzer
  Science Center (Program ID:30471, PI: J.D. Smith). This program is aimed to investigate the
  nature of PAH emission features in nearby galaxies hosting low-luminosity AGN. The observations were obtained
  in low resolution spectral mapping mode with a spectral resolution
  $R\sim60-130$, using the Short-Low ($5-15\mu m$) and Long-Low ($14-38\mu
  m$) modules and taken on February 09, 2007.

  The SL spectral mapping covers $1\times10$ map positions with a
  step size of $1.85''$ in perpendicular direction. Observations
  consist of 60s-ramp duration with 3 circles per step. The observations obtained
  by LL module consist of $4\times11$ map
  positions with a step size of $5.08''$ in the perpendicular direction
  and a step size of $15.0''$ in the parallel direction, 30s-ramp
  duration was used with 1 circle per step. The mapping area is
  shown in Figure 1.

  Primary data reduction were done by Spitzer Science Center (SSC)
  pipeline, version S18.7.0 for both SL and LL modules, including
  standard reductions such as removing of electron bias, dark current
  substraction and flat-field correction. We further use the IDL-based software
  package CUBISM (Smith et al. 2007a) to extract and combine spectra
  into a single spectral cube. Background substraction was done by using off-source observations (rouge)
  accompanied with each map, which were taken in a region containing no spectral signatures by the same slit
  pattern and exposure time with the mapping sequence. After
  background substraction, the major artifacts remaining in the
  two-dimensional map of the data cube are some bright ``stripes'',
  which are caused by bad pixels rastered across the map. The bad
  pixels are removed by carefully examining each wavelength of the panel with the backtracking
  tool in the CUBISM.

\begin{figure}
\includegraphics[width=9cm]{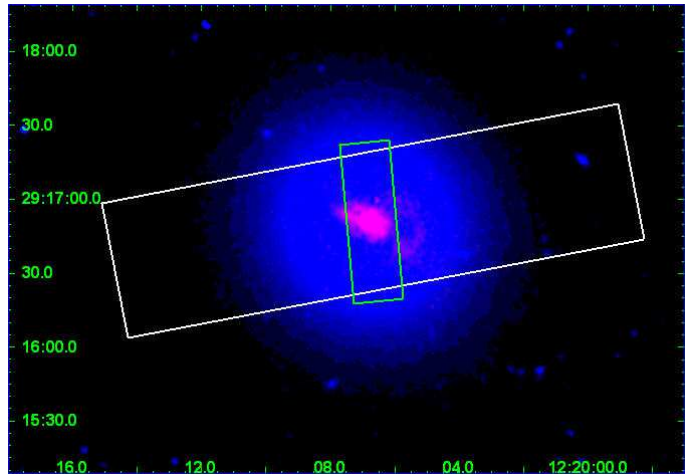}
\caption{Mapping area of NGC 4278. Blue: IRAC3.6 image. Red: IRAC8.0
non-stellar emission, derived by subtracting a scaled IRAC3.6 image
from IRAC8.0 image (Tang et al. 2009). White box: Mapping coverage
of long slit observations. Green box: Mapping coverage of short slit
observations.}\label{fig01}
\end{figure}

\section{Emission Line Distribution}

Ionized gas in NGC 4278 has been previously studied by Sarzi et al.
(2006, 2010) based on the observations of SAURON survey. Figure 2
shows the distribution of [OIII]$\lambda5007$ emission obtained by
SAURON observations (kindly provided by Marc Sarzi) overplotted with
the contours of IRAC $8\mu m$ excess non-stellar emission, the
latter is obtained by subtracting a scaled IRAC 3.6$\mu m$ image
from the IRAC $8\mu m$ image. In Tang et al. (2009), We've shown that 
for quiescent ellipticals, [3.6] - [8.0] color are generally
constant and change little with radius from the center. 
We can simply multiply 3.6 image by a constant to account 
for the contribution of an old stellar population at 8.0um, then degrade this 
image to 8.0um and subtract it from the original 8um image. 
For NCG 4278, this constant, which reveals the infrared-color of an old stars, 
is taken as 0.29.  The
central elongated feature along south-west to north-east direction
with a position angle of $70^{\circ}$ generally coincides with
[OIII]$\lambda5007$ distributions. Both [OIII]$\lambda5007$ and
H$_{\beta}$ emission are further spatially and kinematically
consistent with the inner part of HI disk in NGC 4278 (Morganti et
al. 2006), suggesting that dust, cold gas and warm gas simply belong
to different phases of a same gaseous rotational structure.

\begin{figure}
\includegraphics[width=9cm]{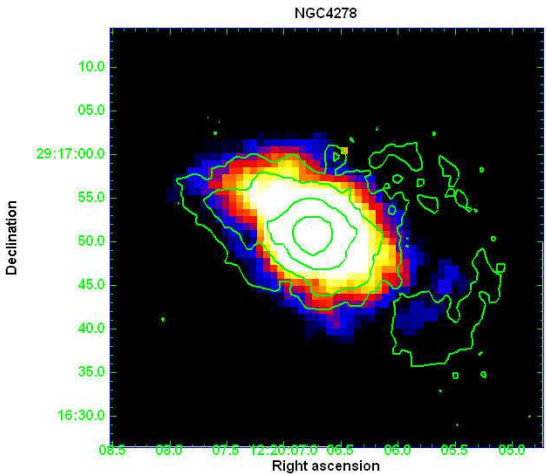}
\caption{[OIII]$\lambda5007$ emission line map, overlaid with
contours of IRAC8.0 non-stellar emission. Contour levels are given
as: 1.50, 0.61, 0.25, 0.10 (in units of MJy/sr).}\label{fig02}
\end{figure}

Another important result of SAURON observations for NGC 4278 is a
``reverse'' distribution of ionization states for warm gas in the
central region. The nuclei showes a lower degree of ionization
comparing with the outer circumnuclear region. In Figure 3 we
present the map of [OIII]$\lambda5007$/H$_{\beta}$ ratio along with
the IRAC $8\mu m$ non-stellar emission. A low-surface brightness,
high ionization ring-like structure extends beyond 10'' and is
obvious seen in the south and west area. This feature, together with
the fact that the radial profile of H$_{\beta}$ flux is more
extended than expected for photoionization by the central AGN (Sarzi
et al. 2010), strongly suggest a second ionization source dominates
over the area outside the central $5''$. The IRAC $8\mu m$
non-stellar emission also shows extended patchy structures to the
west of the main feature. As already noticed by the Sarzi et al.,
the western patchy structures seem to be anti-correlated with the
distribution of ionization degree, with some sub-areas of low
PAH$7.7\mu m$ covering having the highest
[OIII]$\lambda5007$/H$_{\beta}$ ratio. The authors suggested this
feature could be explained by the local hardening of ionizing
continuum and the decline of ionization parameter due to dust
effect. It is interesting to notice that, there is no such area with
irregular [OIII]$\lambda5007$/$H_{beta}$ distribution on the
north-east side of the main elongated structure, where the surface
brightness of $8\mu m$ excess is higher than patchy regions.

\begin{figure}
\includegraphics[width=9cm]{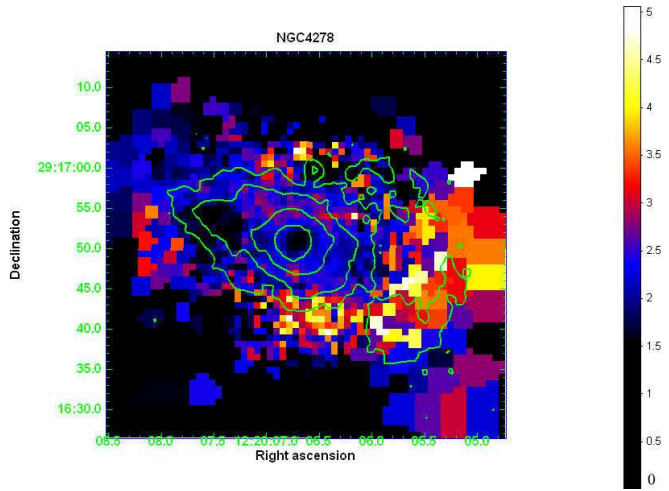}
\caption{Line ratio map of [OIII]$\lambda5007$/H$_{\beta}$, overlaid
with contours of IRAC8.0 non-stellar emission. Contour levels are as
in Fig 2.: 1.50, 0.61, 0.25, 0.10 (in units of
MJy/sr)}.\label{fig03}
\end{figure}

We construct line maps from Spitzer mapping observations by applying standard spectral 
decomposition method (PAHFIT) on each pixel. Deltails on PAHFIT will be discussed in Section 4.
We also compare each emission line map with IRAC $8\mu m$ non-stellar
emission. To do this, we first degrade the resolution of IRAC 8.0 image to that of each
line map by gaussian convolution. The scale of the gaussian keneral is determined by
matching PSF of IRAC 8.0 image with PSF at the wavelength of line center. After convolution, IRAC $8\mu m$ image is rebinned to the same
pixel size of each line map.

In Figure 4 and Figure 5 we show distributions of PAH$11.3\mu m$ and PAH$17\mu m$ band emission respectively,
overlaid on IRAC $8\mu m$ non-stellar emission. For PAH$11.3\mu m$, the mapping area has
not covered the entire PAH emission region. Nevertheless, the two
distributions agree well with each other for the central elongated
feature, supporting the validity of using latter distribution to
trace PAH$7.7\mu m$ emission. The spacial distribution of PAH17, however, does not totally coincide with that 
of IRAC $8\mu m$ excess. PAH$17\mu m$ distributes in a more symmetrical patter relative to IRAC $8\mu m$ excess.

\begin{figure}
\includegraphics[width=6cm]{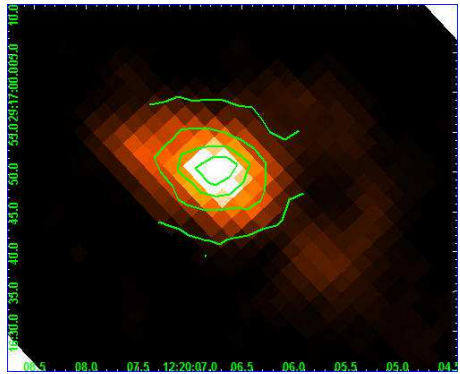}
\caption{IRAC 8.0 non-stellar emission image, overlaid with contours
of PAH11.3 contours. Contour levels are given as:12.1, 8.54, 4.93, 1.33 (in units of $10^{-22}$erg/s), where
the lowest contour level corresponds to $3\sigma$ significance.}\label{fig04}
\end{figure}

\begin{figure}
\includegraphics[width=6cm]{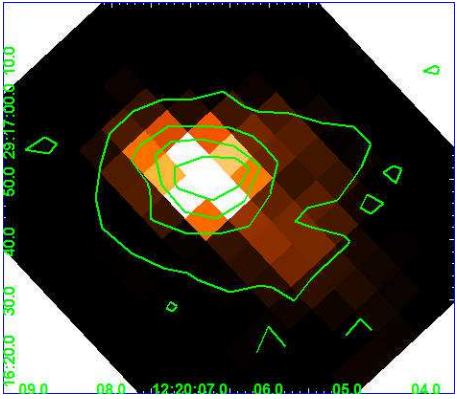}
\caption{IRAC 8.0 non-stellar emission image, overlaid with contours
of PAH17 contours. Contour levels are given as:2.41, 1.67, 0.94, 0.20 (in units of $10^{-21}$erg/s), where
the lowest contour level corresponds to $3\sigma$ significance.}\label{fig05}
\end{figure}

In Figure 6 and Figure 7 we present the distribution of H$_{2}$ S(3)
(mapped in SL module)and H$_{2}$ S(1)(mapped in LL module) pure
rotational lines of molecular hydrogen. For S(3), the distortion of
contours on the northern edge of the main feature results from
removal of a series of severe global bad pixels and does not
necessarily imply a true turbulence. Still, it is obvious that both
S(3) and S(1) emissions generally follows PAH emission. 
The H$_{2}$ S(1) distribution is slighty more exended than PAH patchy
structures west of the main feature. We notice that the flux peak of
H$_{2}$ S(1) map deviates from that of IRAC$8\mu m$ image by about
4'', which is slightly smaller than the pixel size (5.08'') of LL
orders and need to be further confirmed.

\begin{figure}
\includegraphics[width=6cm]{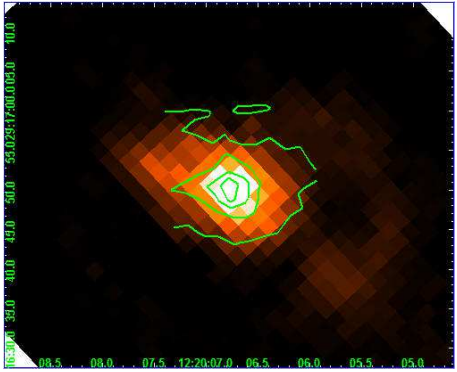}
\caption{IRAC 8.0 non-stellar emission image, overlaid with contours
of H$_{2}$ S(3) contours. Contour levels are given as:4.73, 4.00,
3.31, 2.33, 1.81, 1.36, 0.38 (in units of $10^{-22}$erg/s), where
the lowest contour level corresponds to $3\sigma$ significance.} \label{fig06}
\end{figure}

\begin{figure}
\includegraphics[width=6cm]{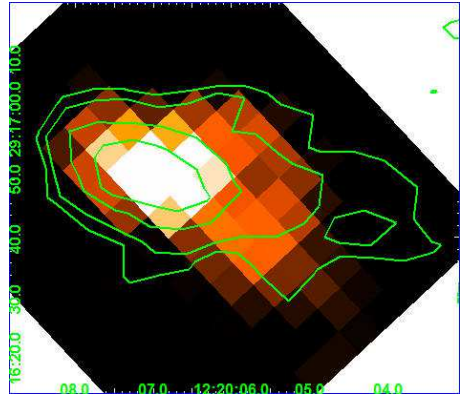}
\caption{IRAC 8.0 non-stellar emission image, overlaid with contours
of H$_{2}$ S(1) contours. Contour levels are given as:4.10, 2.22,
1.09, 0.72 (in units of $10^{-22}$erg/s), where
the lowest contour level corresponds to $3\sigma$ significance.} \label{fig07}
\end{figure}

Figure 8-10 show distributions of fine-structure lines for
[NeII]$12.8\mu m$, [NeIII]$15.6\mu m$ and [SiII]$34.8\mu m$. With
their ionization potential spreading over a wide range, ionized
emission lines show considerable difference from PAH features and
H$_{2}$ pure rotational lines. Especially, for high ionization line
[NeIII] (with an ionization potential of $41.0$ eV), the
central distributions are more symmetrical within the central 10'' comparing
with PAHs and warm molecular hydrogen emissions, although still
showing an faint elongated feature consistent with the position
angle of the latter two distributions. For the strong [SiII]
emission (with an lower ionization potential of $8.2$ eV), its
sufficiently high S/N ratio reveals a more extended component, which
departs from the extended patchy distribution of PAHs and H${_2}$
S(1) emission.

\begin{figure}
\includegraphics[width=6cm]{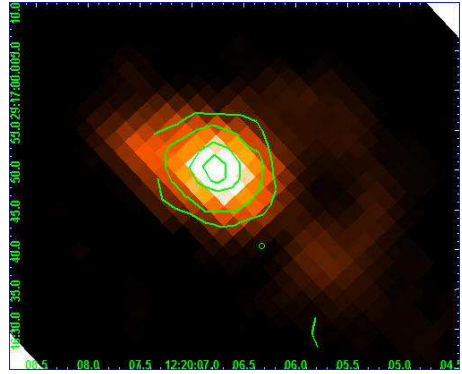}
\caption{IRAC 8.0 non-stellar emission image, overlaid with contours
of [NeII]$12.8\mu m$ contours. Contour levels are given as:3.45,
1.70, 0.64, 0.29 (in units of $10^{-22}$erg/s)., where
the lowest contour level corresponds to $3\sigma$ significance.} \label{fig08}
\end{figure}

\begin{figure}
\includegraphics[width=6cm]{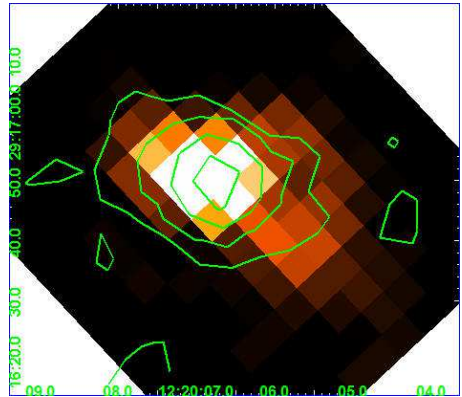}
\caption{IRAC 8.0 non-stellar emission image, overlaid with contours
of [NeIII]$15.6\mu m$ contours. Contour levels are given as:6.34,
3.20, 1.31, 0.68 (in units of $10^{-22}$erg/s), where
the lowest contour level corresponds to $3\sigma$ significance.} \label{fig09}
\end{figure}

\begin{figure}
\includegraphics[width=6cm]{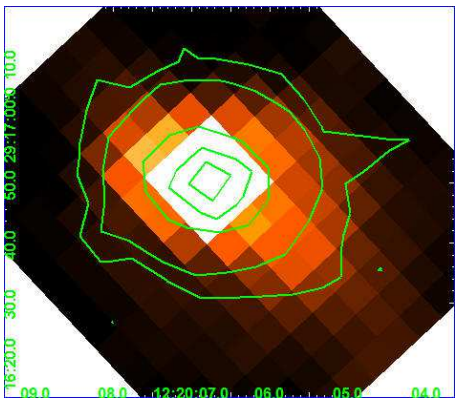}
\caption{IRAC 8.0 non-stellar emission image, overlaid with contours
of [SiII]$34.8\mu m$ contours. Contour levels are given as:16.60, 9.71, 4.80, 1.85, 0.87
 (in units of $10^{-22}$erg/s), where
the lowest contour level corresponds to $3\sigma$ significance.}\label{fig10}
\end{figure}

In Figure 11, we present Chandra X-ray observations of the diffuse
soft X-ray emission with point source being removed(which will be
present in Zhang et al. in preparation), overlaid with the contours
of IRAC 3.6 emission as well as the IRAC 8 non-stellar emission.
While lacking an extended X-ray halo, the bulk of diffuse emission
is confined to the central region and is generally consistent with
the distribution of IRAC 8.0 non-stellar emission. The elongated
feature of X-ray diffuse emission has previously been noticed by
Terashima et al. (2003), this coincidence with warm gas might
suggest a connection between the two phases of gaseous matter in the
center. We also notice that recently, Younes et al. (2010) find a
temperature of $0.4\pm0.1$keV for thermal component in the central
10'' region.

\begin{figure}
\includegraphics[width=8cm]{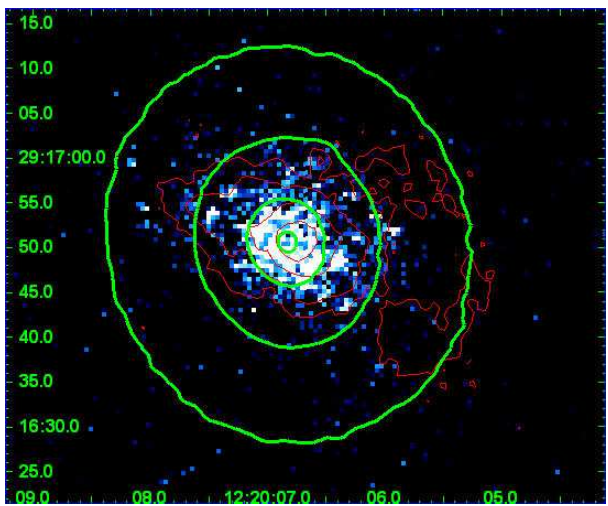}
\caption{Chandra 0.2-1.5keV diffuse emission image removed from
point sources, overlaid with contours of IRAC3.6 image (Green) as
well as contours of IRAC8.0 non-stellar emission (Red). Contours
levels of IRAC3.6 image are given as: 23.22, 6.46, 1.80, 0.50 (in
units of MJy/sr). Contour levels of IRAC8.0 non-stellar emission are
as in Fig 2.: 1.50, 0.61, 0.25, 0.10 (in units of MJy/sr).
}\label{fig11}
\end{figure}

\section{Spectra of Nuclear and Extended Regions}

Spectra of the nuclear region are extracted from a 8-arcsec circular
aperture centered at the surface brightness peak of the continuum.
The mid-infrared emission in NGC 4278 is extended, we need to use
slit loss correction function (SLCF here after) to correct the
original flux calibration provided by SSC based on point-source,
which overestimates the flux by neglecting the flux accepted from
outside the slit. Currently the SLCF is built on a simple assumption
that the extended emission is uniform and unlimitedly extended
(Smith et al. 2007a). Unfortunately, this assumption is not well
satisfied in the case of NGC 4278, which shows a considerable degree
of surface brightness concentration. The application of the SLCF
leads to two artifacts in the spectra. First, a significant mismatch
appears at the boundary of SL2 and LL1, where flux density of SL1 is
$20\%$ higher than LL2. This mismatch happens because the slit width
and pixel size in LL module is nearly threes times as in SL. At the
same wavelength, the SLCF would overcorrect LL spectra in comparison
with SL, because the real ``excess'' flux coming from outside the
slit is relatively less in LL due to concentration of the extended
emission. Another artifact would be more complex. Since the SLCF
depends on PSF sizes at different wavelengths, application of this
correction would modify the spectral profile. We estimate these two
effects by following steps. First, we directly match LL to SL by
multiplication, this spectrum will be referred as "Matched Spectrum"
hereby. Second, we disable the SLCF for LL spectra, this would
elevate LL spectra by about $17\%$ at the blue end, thus fill the
gap between SL and LL, then we do a minor adjustment to scale LL to
SL by multiplication. The resulted spectrum is called "Uncorrected
Spectrum" hereby.The two spectra are shown in Figure 12. The
disagreement between the two spectra is lower than $10\%$ blueward
of $20\mu m$ but becomes significant at longer wavelengths. In this
work, three lines of interest, H$_{2}$ S(0), [SIII]$33.5\mu m$ and
[SiII]$34.8\mu m$, fall on wavelengths longer than $20\mu m$,
discussions depending on relative strengths involved with these
lines will be treated with caution.

We further compared the spectra with IRAC and MIPS photometry
results derived from a same aperture. For the data reduction of IRAC
data, we started from the BCD images and followed the same steps as
we did in Tang et al. (2009), for MIPS data, we started from the
post-BCD image. We employed the extended source aperture correction
for calibration developed by Tom
Jarrett\footnote{http://spider.ipac.caltech.edu/staff/jarrett/irac/calibration/}.
For MIPS image, unfortunately, such a correction is not available.
Thus we only give an estimation of lower and upper limit from the
data. The lower limit is derived by directly integrating the counts
within the aperture without an aperture correction, while the upper
limit is derived by employing an aperture correction for point
source, as provided in MIPS INSTRUMENT HAND
BOOK\footnote{http://ssc.spitzer.caltech.edu/mips/mipsinstrumenthandbook/},
since such an aperture correction ignores the diffraction light from
outside the aperture, it would over-estimate the real flux. The
results are shown in Figure 12. The IRAC photometry points are
consistent with our spectra within calibration uncertainty ($\sim
10\%$, Fazio et al. 2004)  , indicating the flux calibration based
on SLCF is reliable for SL observations. The MIPS photometry data
fails to help distinguish between the Matched Spectrum and the
Uncorrected Spectrum.

Hereafter, we will carry out discussions based on results from the
Uncorrected Spectrum, since the mismatch between the SL and LL
modules indicates that the SLCF correction based on the assumption
of uniform extended source is not proper for LL observations. We
caution the readers that this approach is not a precise solution.
Nevertheless, this will not bring significant influence on the
spectral shape shortward of $20\mu m$, although might overestimate
lines flux by no larger than $30\%$ at the longest wavelength (For
[SiII]$34.8\mu m$, the line flux derived from the Uncorrected
Spectrum is $29.3\%$ higher than that from the Matched Spectrum.).

\begin{figure}
\includegraphics[width=9cm]{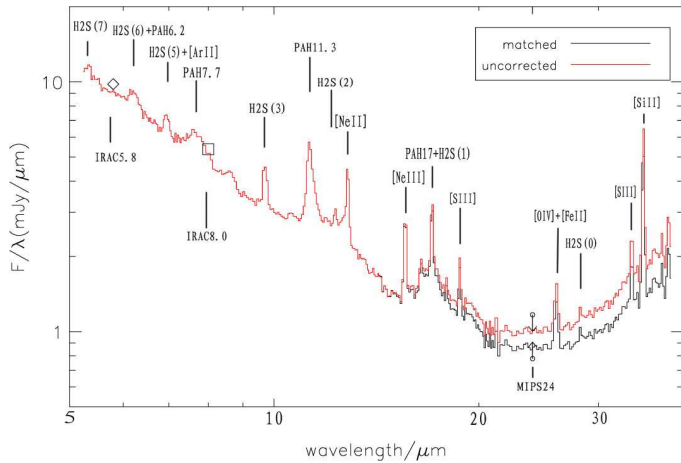}
\caption{Spectra extracted from the central 8'' aperture,
overplotted with IRAC5.8(diamond), IRAC8.0(square) and MIPS24(circle
plus arrow) photometry data points. Black sold line is the Matched
Spectrum with LL orders corrected by SLCF, red sold line is the
Uncorrected Spectrum with LL orders uncorrected by
SLCF.}\label{fig12}
\end{figure}


\begin{table*}
\begin{center}
\caption{Emission Line Strengths in the Central 8'' Aperture
\label{table1}}


\begin{tabular}{lll}
\hline
Line &  ${Flux(m)}^a$ & ${Flux(uc)}^b$   \\
  & $(10^{-21}W/cm^2)$ & $(10^{-21}W/cm^2)$    \\
\hline

{\rm PAH}$6.2\mu m$   &   $22.60 \pm 0.74$    &     $22.50 \pm 0.74$      \\
{\rm PAH}$7.7\mu m$   &   $41.30 \pm 2.46$     &     $40.80 \pm 2.47$      \\
{\rm PAH}$11.3\mu m$  &   $38.60 \pm 0.24$     &     $38.60 \pm 0.24$      \\
{\rm PAH}$12.6\mu m$  &   $17.20 \pm 0.48$     &     $17.00 \pm 0.49$      \\
{\rm PAH}$17\mu m$  &     $22.00 \pm 0.39$     &     $23.80 \pm 0.48$      \\

\hline

H$_{2}$ S(0) & $0.40 \pm 0.04$  & $0.48 \pm 0.06$ \\
H$_{2}$ S(1) & $4.27 \pm 0.10$  & $4.48 \pm 0.13$ \\
H$_{2}$ S(2) & $1.56 \pm 0.07$  & $1.56 \pm 0.07$ \\
H$_{2}$ S(3) & $6.26 \pm 0.09$  & $6.26 \pm 0.09$ \\
H$_{2}$ S(5) & $3.53 \pm 0.27$  & $3.53 \pm 0.27$ \\
H$_{2}$ S(6) & $1.80 \pm 0.22$  & $1.80 \pm 0.22$   \\
H$_{2}$ S(7) & $0.83 \pm 0.24$  & $0.84 \pm 0.24$   \\

\hline

{\rm [Ar II]}$7.0\mu m$  &   $2.78 \pm 0.26$   &  $2.78 \pm 0.26$ \\
{\rm [S IV]}$10.5\mu m$  &   $0.62 \pm 0.09$   &  $0.62 \pm 0.09$ \\
{\rm [Ne II]}$12.8\mu m$ &   $6.14 \pm 0.09$   &  $6.13 \pm 0.09$ \\
{\rm [Ne III]}$15.6\mu m$ &  $4.82 \pm 0.11$   &  $4.95 \pm 0.14$ \\
{\rm [S III]}$18.7\mu m$ &   $1.50 \pm 0.07$   &  $1.65 \pm 0.09$ \\
{\rm [S III]}$33.5\mu m$ &   $1.88 \pm 0.07$   &  $2.39 \pm 0.10$ \\
{\rm [SiII]}$34.8\mu m$ &    $11.00 \pm 0.09$  &  $14.10 \pm 0.15$ \\

\hline

\end{tabular}
\end{center}
$^a$ Spectra with LL orders corrected by SLCF and matched to SL
orders.

$^b$ Spectra with LL orders uncorrected by SLCF.
\end{table*}

We use the package PAHFIT (Smith et al. 2007b) to fit the full
$5-38\mu m$ spectrum. PAHFIT applies three sets of components (thermal
continuum produced by stars and dust; PAH features; emission lines
of ion species and molecular hydrogen) to separate different
emission components. Throughout this article, the error bars are directly given by CUBISM and PAHFIT. 
In CUBISM, only BCD-level statistical uncertainty estimates produced by the IRS pipeline from deviations of the fitted
ramp slope are considered when building error cubes, other uncertainties are 
not taken into consideration(Smith et al. 2008). 
The errors for combined quantities, such as integrated intensities, estimated are formulated 
using the full covariance matrix by CUBISM (Smith et al. 2007b), which is based on Levenberg-Marquardt algorithm.
It should be noticed that systematic calibration uncertainty is not accounted for here. 
We estimate it to be not larger than $10\%$ by comparing SL spectra with IRAC $5.8\micron$ and $8.0\micron$ monochromatic 
flux density (Figure 12). However, in this research, most discussions are based on relative 
strength of spectral features , where calibration uncertainty cancels out.
The result of spectral decomposition is shown
in Figure 13. Generally, the observation could be successfully
reproduced by combination of various components. The spectrum shows
no sign of silicate absorption around $10\mu m$, coinciding with the
result derived from Balmer Decrement (Ho et al. 1997). The only
significant spectral feature unable to be constructed by models is
the $18-20\mu m$ plateau possibly caused by large PAH molecules,
simply because PAHFIT excludes this feature from consideration.
Smith et al. (2007b) suggested this plateau-like extension is not
correlated with PAH $17\mu m$ feature. Another potential problem
underlying the procedure of PAHFIT is that it only considers
silicate absorption. As pointed out by Bregman et al. (2006) and
Kaneda et al. (2008), for early type galaxies, silicate emission
feature around $10\mu m$ originated from the circumstellar envelope
of asymptotic giant branch (AGB) stars might affect the measurement
of PAH feature at $7.7\mu m$. This effect will be estimated in the
next section.

\begin{figure}
\includegraphics[width=7cm]{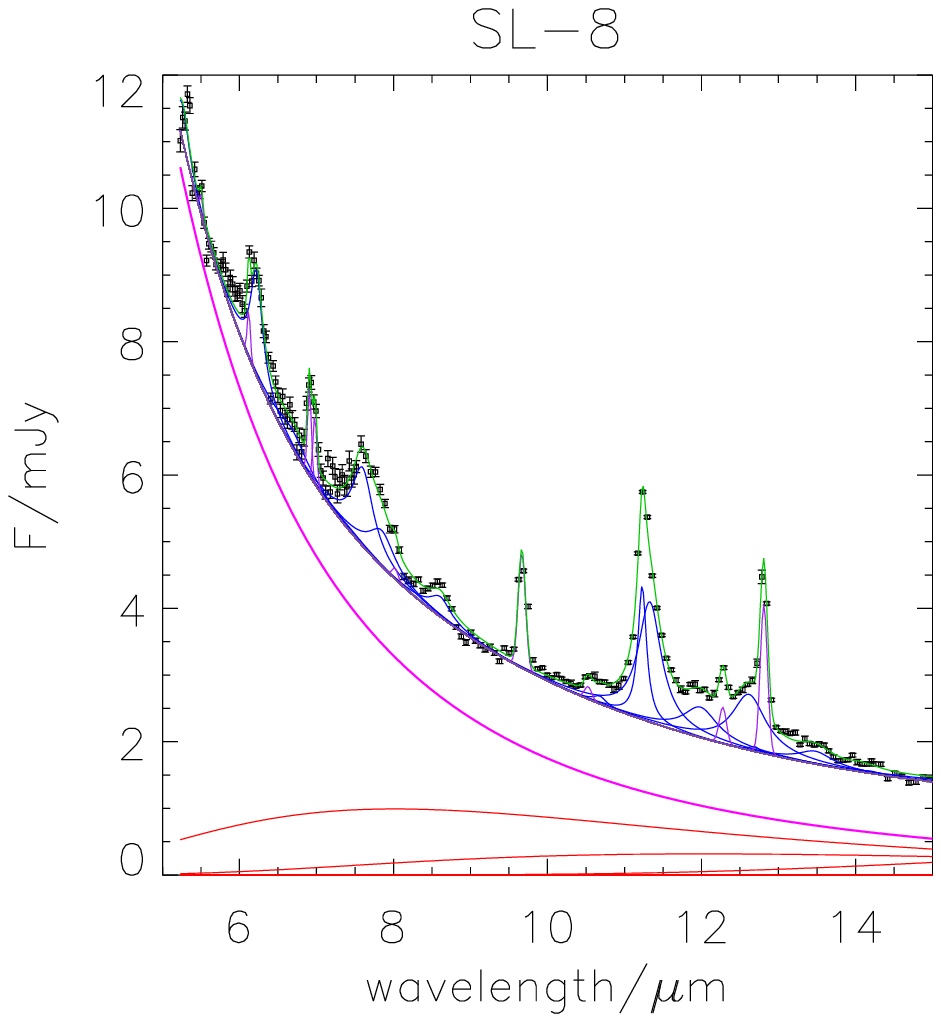}
\includegraphics[width=7cm]{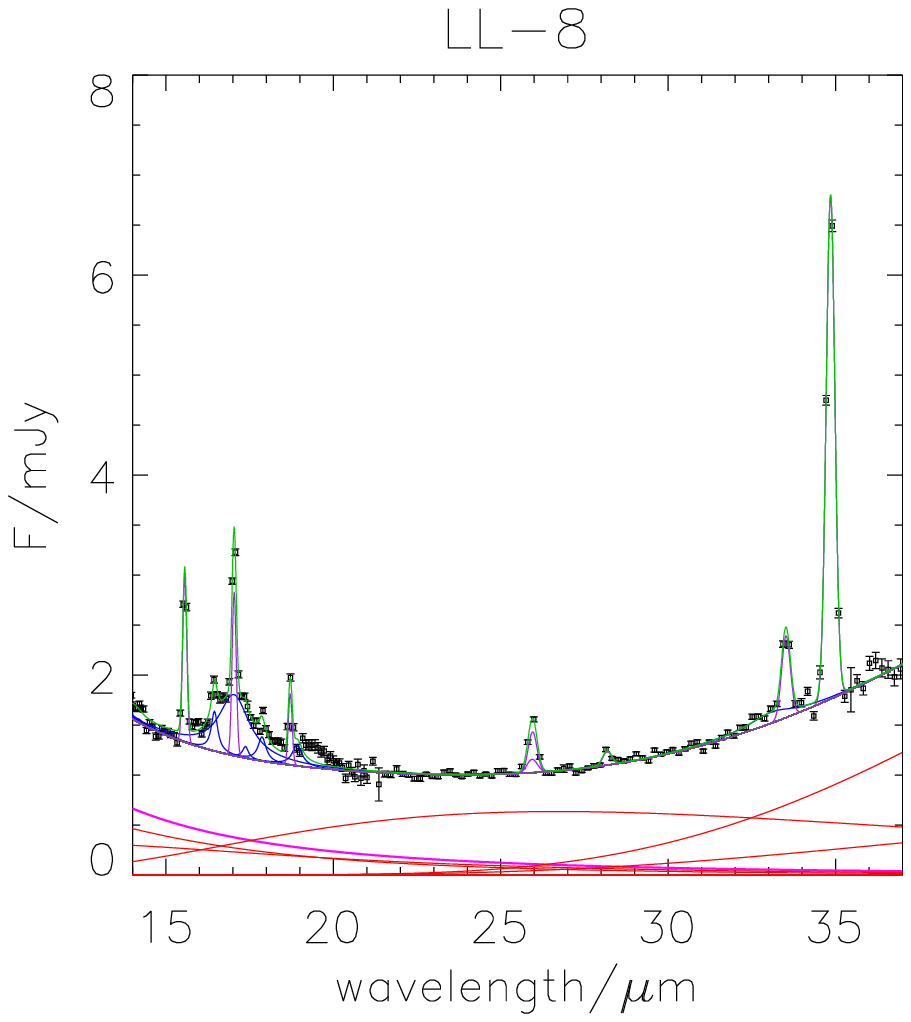}
\caption{Full decomposition of Uncorrected Spectra extracted from
the central 8'' aperture. The magenta line shows the stellar
continuum, red solid lines represent the multi-temperature thermal
dust components, the grey line is the summation of the stellar and
dust continuum. Blue lines show PAH emission features, violet lines
represent emissions arising from warm molecular hydrogen and ionized
species. Up: $5-15\mu m$, Down: $15-37\mu m$.}\label{fig13}
\end{figure}

The nuclear spectra of NGC 4278 are somehow beyond the ordinary
elliptical galaxies for its abundance of emission features. In
contrast, the continuum generally follows a blackbody radiation from
an old stellar population, with some marginal contribution from hot
and warm dust. The emission features appearing in the mid-infrared
spectra of NGC 4278 could be divided into three groups, pure
rotational emission lines of molecular hydrogen, forbidden emission
lines of ions and PAH emission features. In the following sections,
we extract physical information from each set of lines, then we will
discuss possible energy origin of emission features implied by
properties of the gaseous matter.

\subsection{PAH Emission Features}

Detection of PAH emissions is uncommon for elliptical galaxies,
which could naturally be explained by rapid destruction due to
sputtering by hot plasma. Nevertheless, as mentioned before, NGC
4278 is devoid of a X-ray halo. While strong PAH emissions are
always related with star-forming activities (Wu et al. 2005), this
might not necessarily be the case in elliptical galaxies, since
significant amount of UV photons could be expected for an old
stellar population (Binette et al. 1994; Greggio \& Renzini et al.
1990). In some limited cases reported by now, such as Bregman et al.
(2006a) and Kaneda et al. (2008), PAH emission features in
ellipticals show obvious difference from galaxies with active
star-formation. The band ratio of $7.7\mu m$ emission feature
arising from CC stretching vibrations of PAHs , to $11.3\mu m$
feature arising from CH out-of-plane bending vibrations could be
lower significantly than star-forming regions. This is also the case
for NGC 4278, with ${I_{7.7}}/{I_{11.3}}=1.07$. Kaneda et al. (2008)
pointed out that while using PAHFIT to derive PAH emission flux,
silicate emission feature around $10\mu m$ originated from outflow
winds of AGB stars could lead to significant underestimation for PAH
7.7 feature. When this feature is subtracted from the original spectrum of NGC 4278, 
the continuum on the long wavelength side of PAH7.7 will be depressed, 
thus PAH 7.7 flux derived from fitting will increase significantly,
 especially when PAH 7.7 emission is weak. To estimate this effect, 
we subtract a template of
quiescent elliptical galaxy after scaling it to our spectrum at
$5.5\mu m$ (which is an average of three quiescent ellipticals: NGC
1407, NGC 1549 and NGC 3904, kindly provided by Hidehiro Kaneda),
the resulted PAH 7.7 emission doubled in flux and gave
${I^*_{7.7}}/{I_{11.3}}=2.03$. In comparison, the median value of
${I_{7.7}}/{I_{11.3}}$ is about 4 for HII dominated sources in SINGs
sample (Smith et al. 2007b), with a minimum around 2.

In actively star-forming galaxies, PAH band ratio of $7.7/11.3$ has
been suggested to reflect the relative proportion of ionized PAHs to
neutral PAHs (Galliano et al. 2008), which is further controlled by
intensity of UV radiation field, electron density and temperature.
Meanwhile, temperature distribution of PAHs could also affect PAH
band ratios, while higher average temperature generally lead to
stronger emission at shorter wavelength. It is still in argument,
however, which factor controls PAH band ratio in non star-forming
regions, such as elliptical galaxies or AGNs. Smith et al. (2007b)
compare HII sources with AGNs in SINGs sample, they found lower
ratios of ${I_{7.7}}/{I_{11.3}}$ and ${I_{11.3}}/{I_{17}}$ for PAH
emissions in AGNs and also lower ratio of the PAH luminosities to
total infrared luminosities for AGNs. All these facts are consistent
with the scenario that the selective destruction of small PAH
molecules and the resulted shift of temperature distribution toward
lower temperature are responsible for PAH band ratios in AGNs. This
viewpoint is supported by O'Dowd et al. (2009), who used optical
diagnostics to clarify AGNs and star-forming galaxies and found that
the PAH band ratios behave as the destruction of small PAH molecules
becomes important in AGNs. In this sense, NGC 4278 is consistent
with the AGN population in all respects. The band ratio
${I_{11.3}}/{I_{17}}=1.75$, also falls in the AGN region of Smith et
al. 2007. We further estimate the ratio of total PAH luminosity to
total infrared luminosity. Following Dale et al. (2002) and Smith et
al. (2007b), we obtain the total infrared luminosity based on MIPs
observations at $24, 70, 160\mu m$ (data taken from Temi et al.
2009). Then we calibrate the TIR luminosity to $8''$ aperture based
on MIPS 24 luminosity derived from the spectrum. The result gives a
${L_{total PAHs}}/{L_{TIR}}=4.7\%$, which is lower than HII galaxies
in SINGs sample (with a median$\sim 10\%$, and a minimum $\sim 8\%$
for HII galaxies with metallicity $12+log(O/H)>8.1)$.

\subsection{Warm Molecular Hydrogen}

Mid-IR molecular hydrogen quadrupole emission lines arising from
pure rotational transitions are clearly detected in the central
region of NGC 4278. These lines are effective coolants for warm
molecular gas with temperature $\sim 100-1000K$. Excitation
mechanisms of H$_{2}$ emissions include radiative decay after
pumping to electronically excited states through absorption of
ultraviolet photons (Morton \& Dinerstein, 1976); inelastic
collision with other particles(Martin et al. 1996; Everett \& Pogge,
1997); reformation onto excited states for dissociated molecular
gas(Hollenbach \& Mickee, 1989). Due to low critical densities ($<$
a few $\times 10^3 cm^{-3}$ for S(0)-S(3)), transitions between
lower levels are usually thermalized through collisional
de-excitation, making pure rotational lines alone not guaranteed
diagnostics to distinguish between different excitation mechanisms.

The radiative decay within pure rotational levels of molecular
hydrogen follows quadrupole selection rules, only $\bigtriangleup
J=-2$ could occur, where J is the rotational quantum number. The
flux of a certain transition could be expressed as
${F_{ul}}={A_u}{N_u}{h{\nu}_{ul}}{\Omega}/4{\pi}$, where u=l+2,
$A_u$ is the Einstein A coefficient of the upper level, ${N_u}$ is
the column density of molecules on the upper level, $h{{\nu}_{ul}}$
is the photon energy and $\Omega$ is the beam solid angle. Under the
assumption of local thermalization equilibrium (LTE), the fraction
of molecules in each states to total column density could be written
as ${N_u}/{N_{total}}={g_u}{exp[-{E_{ul}}/kT]}/{Z(T)}$. Where $g_u$
is the statistical weight, given by $g_u=(2I+1)(2l+1)$, here I=0 for
even u and I=1 for odd u. Z(T) is the partition function. We adopt
$Z(T)=0.0247T/[1-exp[6000K/T]$ (Herbst et al. 1996), which is valid
for $T>40K$.

Figure 14 shows the excitation diagram for the central 8'' region of
NGC 4278. Emission lines up to S(7) are detected above $3\sigma$
level, except S(4), which could be obscured by PAH7.7 feature.
The fit of S(6) is highly affected by the adjacent
6.2 micron PAH feature. We estimate the uncertainty of S(6) by fitting 100 
randomly noise-added spectra within noise limit, the result estimated in this way 
is in consistent with that directly derived from PAHFIT, with a mean flux of $1.8\times10^{-21}erg/s$.
Data points in our diagram do not show obvious sign of ortho-to-para
ratio (OPR) deviating from thermal equilibrium below $J=8$ level,
which would appear as a ``zigzag'' pattern between ortho states and
para states (Neufeld et al. 2006). We thus use a two temperature
model with $OPR=3$ (equilibrium value) to fit all data points. The
result gives a warm component $T_w \sim 197K$ with with column
density $N_{Hw} =0.75\times10^{20} cm^{-3}$ and a hot component
($T_h \sim 799K$) with ${N_{Hh}}^{hot}=1\times10^{18} cm^{-3}$,
corresponding to a mass of $1.49\times10^6M_{\odot}$ within 8''
aperture. This result remains almost unchanged if we apply lines
derived from the ``Matched Spectrum'', which lowers S(1) by $10\%$
and S(0) by $18\%$, the two temperature given in this case is $T_w
\sim 206K$ and $T_h \sim 811K$.

\begin{figure}
\includegraphics[width=9cm]{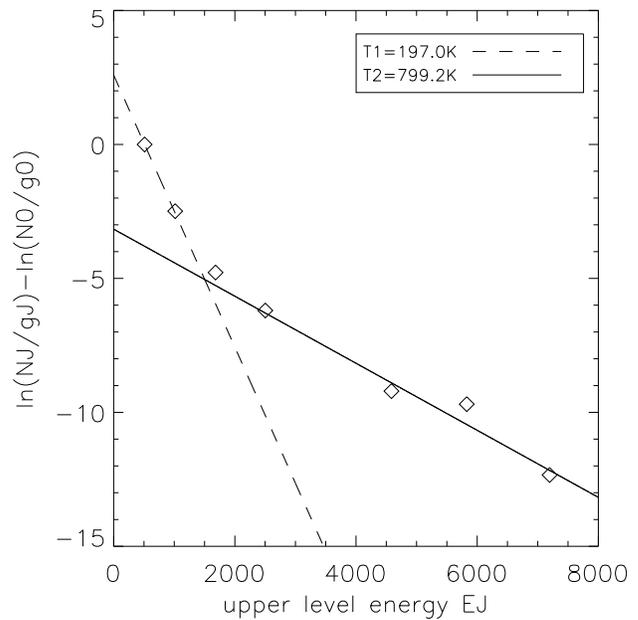}
\caption{Excitation diagram of H$_{2}$ pure rotational lines. The
dash line represents a temperature of 197.0K, the solid line
represents a temperature of 799.2K. The data point of H$_{2}$ S(4)
is not shown, since it is blended with PAH$8.6\mu m$ features and
could not be accurately decomposed.}\label{fig14}
\end{figure}

Roussel et al. (2007) carried out a systematic research on H${_2}$
pure rotational emissions based on SINGs galaxies sample, they
suggest that H${_2}$ rotational lines in nearby star-forming
galaxies are dominated by emission from photon dissociation regions
(PDR), while in AGN objects excess H${_2}$ emissions could be
excited through shock heating. In NGC 4278, molecular hydrogen shows
a higher excitation temperature than most galaxies in SINGs sample,
in which only three galaxies classified as AGN objects are detected
of lines higher than S(3). To compare NGC 4278 with SINGs galaxies,
we follow Roussel et al. (2007) and use only S(0)-S(3) to fit a two
temperature model, which gives a warm component $T_w \sim 189K$ with
$N_{Hw}=2.1\times10^{20} cm^{-3}$ and a hot component $T_h \sim
701K$ with $N_{Hh}=4\times10^{18} cm^{-3}$. The temperature of the
warm component and the fraction of the hot component are still
higher comparing with most SINGs galaxies. Meanwhile, this
temperature is lower than stellar scale shock-heated region, such as
supernova remnants (Neufeld et al. 2007; Hewitt et al. 2009) and
Herbig-Haro objects (Neufeld et al. 2006), but similar to some
recently discovered galaxy-scale shocks(Appleton et al. 2006; Ogle
et al. 2007). Furthermore, CO emissions have been detected in NGC
4278 with IRAM observations (Combes et al. 2007). This allows us to
calculate the fraction of warm H$_{2}$ to cold H$_{2}$. We assume
that the excitation temperature of warm H$_{2}$ keeps constant and
calibrate the mass from 8'' aperture to 23'' aperture based on the
flux of H$_{2}$ S(1), 23'' is the beam size of IRAM to derive cold
H$_{2}$ mass. The result gives M(warm H$_{2}$)/M(cold H$_{2}$)=0.10,
somehow larger than the typical value in SINGs sample.

A three temperature model would not achieve a significantly better
result, this is because energy levels higher than $J=7$ do not lie
along a straight line and could not be fitted by a simple
temperature component. One possible explanation is that S(6) line
(J=8) is overestimated, since this line is blended with  [ArII]
emission and might not be well decomposed. Otherwise, if S(6) and
S(7) are measured accurately, higher rotational levels with $J>7$
might indicate a OPR lower than 3. Unfortunately, without a reliable
measurement of S(4), which is blended with PAH$8.6$ feature, we are
not able to discern between different possibilities.

\subsection{Fine-structure Line emission}
The nuclear spectra of NGC 4278 show abundant emission lines arising
from low-level forbidden transitions of different ionic species,
including [ArII]$7.0\mu m$, [SIV]$10.5\mu m$, [NeII]$12.8\mu m$,
[NeIII]$15.6\mu m$, [OIV]\&[FeII]$26\mu m$ blend, [SIII]$18.7\mu m$,
[SIII]$33.5\mu m$ as well as [SiII]$34.8\mu m$. These lines provide
valuable information about physical states of warm gas. The ratio of
[NeIII]/[NeII] is commonly used as an indicator of ionization
states, which shows a value of 0.81 for the nuclear region of NGC
4278. This is a typical value found in LINER objects and cannot help
distinguish among starforming regions, AGN and shock-heated regions
(Allen et al. 2008, Dale et al. 2009).

The most distinctive feature of the fine-structure emissions in NGC
4278 is the strong [SiII]$34.8\mu m$ emission located at the red end
of the spectrum, which could be compared with the nearby
[SIII]$33.5\mu m$ line and shows a ratio of [SiII]$34.8\mu
m$/[SIII]$33.5\mu m$=5.9. This ratio is higher than any of SINGs
galaxies (Dale et al. 2009) and resemble values observed in
shock-dominated regions, such as supernova remnants (Neufeld et al.
2007) and large scale shock triggered by galaxies interaction
(Cluver et al. 2010), in which [SiII] emission is boosted by
sputtering destruction of dust during fast-shock, since as an
element of high depletion, a large fraction of Si is locked in dust
in diffuse ISM. Dale et al. (2009) find a separation of
[SiII]/[SIII] ratio between AGNs and HII galaxies in SINGs sample
while AGNs show stronger [SiII] emissions. In their discussion, this
separation might be caused by reduced depletion of silicon, X-ray
photoionization process or an increased density for emission gas.
The last possibility, however, seems unlikely to be applicable to
NGC 4278. The electricity density derived by the ratio of
[SIII]$18.7\mu m$/[SIII]$33.5\mu m$, with a value of 0.69, is close
to the low density limit and indicates the bulk of the
fine-structure emissions come from regions with density around $300
cm^{-3}$ (Giveon et al. 2002). On the other hand, it is also
doubtable that X-ray photoionization plays a major role in boosting
[SiII] emission in the nuclear region of NGC 4278, which holds a
LLANG with a moderate hard X-ray luminosity
$L_{2-10keV}=10^{40.4}erg/s$, since as mentioned above, the
ionization of extended gas is not dominated by AGN photoionization,
and [SiII] line distribution shows a widely extended structure, we
will see in Section 4.5 that high [SiII]/[SIII] ratio still appears
in area more than 1kpc away from the nuclei.

\subsection{Radial Emission-Line Distributions}
It is important to understand the excitation mechanism of the
extended emission in the central region. However, results given
above are based on the spectra extracted from the central 8''
aperture and could not avoid being mixed with an AGN component. It
is thus very useful to further quantitatively explore the spatial
distributions of different emission features. Unfortunately, both
the limited mapping area of the SL module as well as the imperfect
calibration for extended emission pose obstacles to this purpose. In
this section, we extract spectra from three adjacent apertures of
5'' radius, with one centered at the continuum peak of the spectral
map and two centered at the northeast and southwest of the galaxy
center, as shown in Figure 15. The 5'' aperture size is somehow
small for extraction of LL orders due to its large pixel size and
extended PSF, comparing radial distribution of different lines based
on LL observations should be taken with caution. Therefore, here we
only focus on SL spectra. The spectra are presented in Figure 16 and
the detected line fluxes are shown in Table 2.

\begin{figure}
\includegraphics[width=6cm]{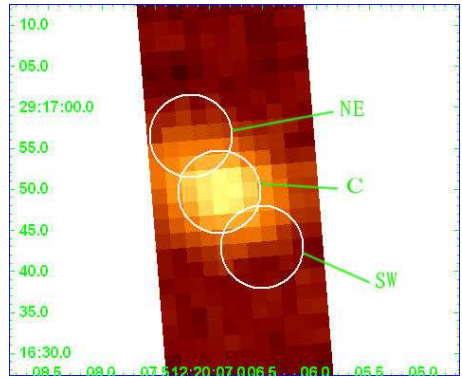}
\caption{PAH11.3 map, overplotted with three neighbouring extraction
apertures, each has a 5'' radius: North-East(NE, centered at
185.02986,29.28235), Central(C, centered at 185.02875,29.28048),
South-West(SW, centered at 185.02802, 29.27872). }\label{fig15}
\end{figure}

While [NeIII]$15.6\mu m$/[NeII]$12.8\mu m$ is a comfortable
indicator of ionization states, [NeIII] falls on the LL wavelength
range. Here we use [SIV]$10.5\mu m$/[NeII]$12.8\mu m$ as a
substitute for diagnostic of gas ionization. The validity of this
approach has been verified by Groves et al. (2008). We find that the
[SIV]/[NeII] in the two extended regions, say, NE and SW regions,
are higher than the central region, although in both cases the [SIV]
fluxes are barely higher than the $3\sigma$ level. The [SIV]/[NeII]
ratio is 0.16 in NE region and 0.24 in the SW region, in comparison
with 0.08 in the central region. This result is consistent with the
distribution of [OIII]$\lambda5007$/H$_{\beta}$ derived from SAURON
observations and confirm that a significant fraction of extended
emission is not excited by central AGN photoionization even within
the central 10'' region.

\begin{figure}
\includegraphics[width=9cm]{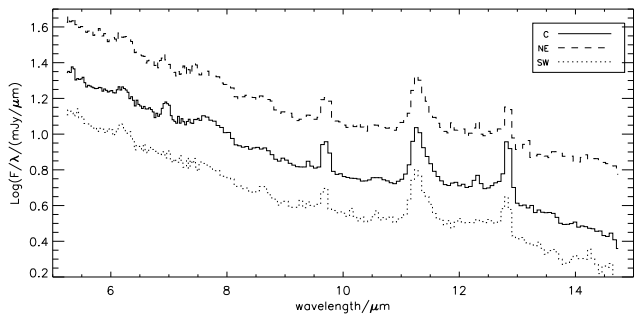}
\caption{Scaled SL spectra extracted from the three apertures shown
in Fig 14. The solid represents the result from the central
aperture, the dash line represents the result from the north-east
aperture, the dot line represents the result from the south-west
aperture.}\label{fig16}
\end{figure}

\begin{figure}
\includegraphics[width=6cm]{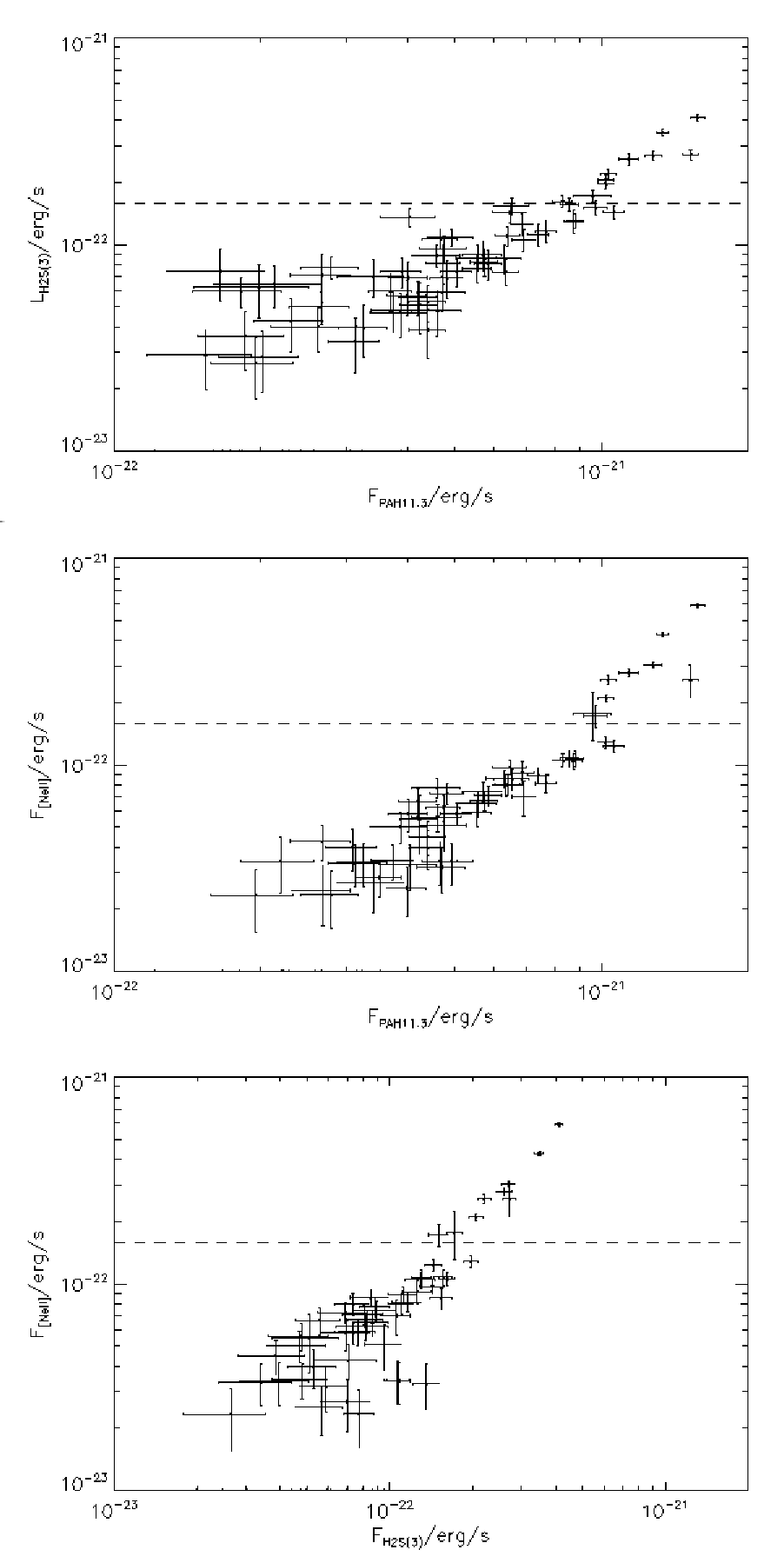}
\caption{Pixel-to-Pixel plots for central $9\times11$ pixels of SL mode spectral map. Upper panel: H$_{2}$ S(3) to PAH11.3. 
Middle panel: [NeII] to PAH11.3. Lower panel: [NeII] to H$_{2}$ S(3). 
Dash line mark pixels where slope starts to change.}\label{fig16}
\end{figure}


\begin{table*}
\begin{center}
\caption{Emission Line Strengths in 3 Neighboring 5'' Apertures
\label{table2}}


\begin{tabular}{llll}
\hline
Line &  NE & C  & SW   \\
  & $(10^{-21}W/cm^2)$ & $(10^{-21}W/cm^2)$ & $(10^{-21}W/cm^2)$   \\
\hline

{\rm PAH}$6.2\mu m$  & $3.63 \pm 0.12$ & $ 8.63 \pm 0.14$ & $2.79 \pm 0.15$ \\
{\rm PAH}$7.7\mu m$  & $16.05 \pm 1.06$ & $ 51.77 \pm 1.10$ & $12.86 \pm 1.03$ \\
{\rm PAH}$11.3\mu m$  & $8.67 \pm 0.17$ & $ 22.10 \pm 0.15$ & $5.96 \pm 0.13$ \\
{\rm PAH}$12.6\mu m$  & $2.83 \pm 0.29$ & $ 9.03 \pm 0.30$ & $3.04 \pm 0.12$ \\
\hline

H$_{2}$ S(2) & $0.33 \pm 0.04$ & $0.98 \pm 0.05$ & $0.15 \pm 0.04$\\
H$_{2}$ S(3) & $1.21 \pm 0.05$ & $4.08 \pm 0.06$ & $0.79 \pm 0.06$\\
H$_{2}$ S(5) & $0.74 \pm 0.15$ & $2.46 \pm 0.18$ & $<0.34$\\
H$_{2}$ S(6) & $0.44 \pm  0.14$ & $1.15 \pm 0.14$ & $0.49 \pm 0.14$\\
H$_{2}$ S(7) & $<0.43$  & $0.75 \pm 0.15$ & $<0.46$\\

\hline
{\rm [ArII]}$7.0\mu m$ & $<0.54$ & $2.34 \pm 0.17$ & $<0.36$\\
{\rm [SIV]}$10.5\mu m$ & $0.16 \pm 0.05$ & $0.33 \pm 0.06$ & $0.16 \pm 0.05$\\
{\rm [NeII]}$12.8\mu m$ & $1.00 \pm 0.04$ & $ 4.46 \pm 0.06$ & $0.67 \pm 0.03$\\
\hline
$9.0\mu m ^a$ & $7.85 \pm 0.21$ & $20.40\pm 0.23$ & $6.01 \pm0.18$\\
\hline
\end{tabular}
\end{center}
$^a$ 9.0 Continuum Flux Densities are given in units of mJy.
\end{table*}

It is also intersting to explore any possible variation of PAH band ratio in three regions.
However, we found that PAH6.2/PAH7.7 ratios directly derived from PAHFIT in three regions are larger than 0.5 and cannot
be explained by any PAH model (Draine \& Li, 2001). We suggest such an
unphysical value of PAH6.2/7.7 appeared because we failed to consider the silicate emission as we did in Section 4.1, 
thus underestimated PAH7.7 flux.
With a template of silicate emission being subtracted from original spectra, the fitting results are given in Table 2.  
Our results show that both PAH6.2/7.7 and PAH7.7/11.3 do not have significant spatial variation, 
while PAH 6.2/7.7 is slightly lower in the center and
PAH7.7/11.3 is higher in the center. This may suggest more large, neutral PAHs in the galaxy center relative to the outer area,
which could be a result of either hard radiation process 
by AGN or shock process while gas flow into the center.

In Figure 17, we show pixel-to-pixel comparisons between [NeII]$12.8\mu m$, PAH$11.3\mu m$
and H$_{2}$ S(3) for central $11\times9$ pixels of SL map. 
It is obvious that [NeII]$12.8\mu m$ shows distinct slope changes for most luminous 9 pixels, 
which correspond to a central region of about $<3''$. 
This should be a sign that AGN continuum starts to dominate the heating in this 
region and produces relative more high ioniztion nebula line emissions relative to PAHs emissions 
and warm molecular emissions. This result confirms
SAURON result that [OIII]$\lambda5007$/H$_{\beta}$ starts to decline as entering 
central 5'' area. Another sign revealed by Figure 17 is 
enhanced H$_{2}$ S(3) emission toward the outmost area, relative to both [NeII] 
and PAH11.3, which appears as a shallower slope in H$_{2}$ S(3) to PAH$11.3\mu m$ plot 
and excessive H$_{2}$ S(3) in [NeII]$12.8\mu m$ to H$_{2}$ S(3) plot. Enhanced
H$_{2}$ S(3) toward the outter area is consistent with our finding that H$_{2}$ S(1) is
more extended than both PAH emissions and nebula lines.

\subsection{Spectra in Western High-Ionization Region}

\begin{figure}
\includegraphics[width=6cm]{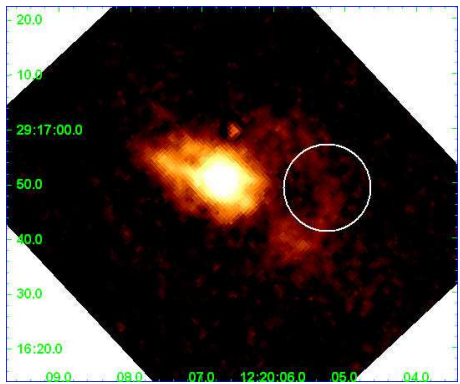}
\caption{Image of IRAC8.0 non-stellar emission, overplotted with the
extraction aperture for high ionization region, which is centered at
(185.02190,29.28036), with a radius of 8''.}\label{fig18}
\end{figure}

We further explore the ionization and excitation states for gas in
the western high-ionization, low surface brightness region as
revealed by SAURON observation. We extract LL spectra from a 8''
aperture centered at the west of the nuclei, as shown in Figure 16.
Unlike the nuclear region, here we employ SLCF to correct spectra
since there should be considerable outcoming light from the nearby
bright nuclear region diffracted into local area. The spectrum is
given in Figure 17, comparing with spectrum extracted from the
central area. For high-ionization region, the continuum emission
longer than $20\mu m$ show a shallower slope than spectra of the
central region, this difference remains when we compare it with the
SLCF-corrected central spectrum, which, as discussed above, flattens
the continuum longer than $20\mu m$. We can quantify the slope longer than 20um continuum 
simply by the ratio of final fitted continnum at 36um to that at 21um. 
For central region with slit loss correction, F36/F21=2.7, for central region 
without SLCF, F36/F21=3.1, while in high ionization region, the ratio is 1.9.  The lack of a hot dust
component, as indicated by shallower continuum could be a natural
result of weaker UV-optical continuum or larger sizes of local dust
grain.

The high-ionization region is characterized by an enhanced H$_{2}$
S(1) emission. The relative strength of H$_{2}$ S(1) to nearby
[NeIII]$15.6\mu m$, [SIII]$18.7\mu m$ ionized emission and
underlying PAH$17\mu m$ feature in high-ionization region is
increased by a factor of 2 or more. Meanwhile, the excitation
temperature derived from H$_{2}$ S(1) and H$_{2}$ S(0) is $>183K$.
Without information from higher level lines, this result indicates a
temperature not lower than the central region if assuming a
equilibrium OPR. The enhancement of H$_{2}$ S(1) is consistent with
what has been shown in Figure 6, the distribution of H$_{2}$ S(1)
narrow band emission exhibits an extended component to the west of
the nuclei, in the next section, we will discuss possible origin of
this feature.

\begin{figure}
\includegraphics[width=9cm]{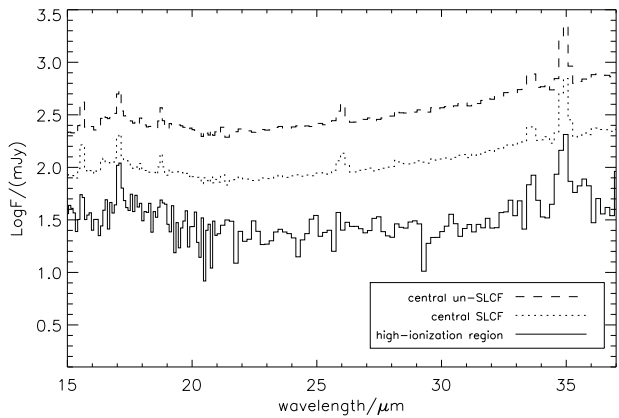}
\caption{Scaled LL spectra (solid) extracted from the aperture shown
in Fig 17. For comparison, Central 8'' Matched Spectra (dot line)
and Uncorrected Spectral (dash line) are scaled and plotted.
}\label{fig19}
\end{figure}


\begin{table}
\begin{center}
\caption{Emission Line Strengths in the High-Ionization Region
\label{table3}}


\begin{tabular}{ll}
\hline
Line &  Flux  \\
  & $(10^{-21}W/cm^2)$   \\
\hline

{\rm PAH}$17\mu m$  & $2.24 \pm 0.38$\\
\hline

H$_{2}$ S(0) & $<0.12$  \\
H$_{2}$ S(1) & $0.85 \pm 0.07$   \\

\hline
{\rm [NeIII]}$15.6\mu m$ & $0.30 \pm 0.06$\\
{\rm [SIII]}$18.7\mu m$ & $<0.19$ \\
{\rm [SIII]}$33.5\mu m$ & $0.26 \pm 0.06$\\
{\rm [SiII]}$34.8\mu m$ & $0.95 \pm 0.10$\\
\hline

\end{tabular}
\end{center}
\end{table}

\section{Discussion: The Origin of Extended Warm gas and Dust}

Results from Spitzer mapping observations reveal abundant warm dust,
molecular gas and ionized gas in the central region of NGC 4278. One or more additional
ionization sources apart from AGN photoionization are required
to explain the extended ionized gas. It is conceivable that energy source resoponsible for extended
warm gas should also be related to the
physical properties of extended dust and molecular hydrogen we
observed. In this section, we will discuss possible heating
mechanisms that could offer a self-consistent explanation for the
different phase of extended dusty gas.

Both the detections of PAHs and cold molecular gas might imply some
degree of residual starformation, while it seems unlikely that
starforming activity plays an important role in the case of NGC
4278. Infrared emission lines NGC 4278 show distinct behavior 
from that should be expected for PDR region (Kaufman et al. 2006). Besides, barely any sign of starforming activity is shown
at any wavelength in the central kpc region. Galex observations show that NGC 4278 should be 
classified as a typical UV upturn galaxy based on its strong FUV emission throughout the galaxy, this is a direct evidence 
that the background radiation field is possibily from a hotter 
old stellar population (post AGBs, LMXBs or EHBs) instead of OB stars. 
Second, in Sarzi M., Shields, J. C., Schawinski K. et al. (2010), the authors have 
quantitativly explored the ionization mechanism for NGC 4278 based on comparison
 between emission line diagnostic and predictions of MAPPING-III models (Section 3.4.1, Figure 12). 
Their results show that ionized gas in NGC 4278 clearly fall outside the starformation region 
in emission line diagram. The extended ionized gas in NGC 4278 exhibits LINER type emission lines, 
which further cast doubt on the idea that ionized gas is not heated by young stars.

As to stellar population analysis, previous research show different results for the 
stellar population of NGC 4278, while Serra P. et al. 2008 reported a 3Gyr single stellar 
population in a circumnuclear region of NGC 4278, Zhang Y. et al. 2008 and recent SAURON results
 (Kuntschner, H. et al. 2010) just gave a ~10Gyr stellar population. 

If the bulk of extended atomic, molecular and dust emission comes from starforming area, 
similar spacial distribution should be expected for different types of emission lines,  
such coinsidence is also not observed, suggesting at least a second physical process should be acting.

Another commonly employed heating mechanism in elliptical galaxies
is process related to X-ray emitting hot gas, such as cooling flow
from hot gas and evaporation flow caused by interaction between cold
and hot gas (Sparks et al. 1989; de Jong et al. 1990). Still, this
possibility is easy to be ruled out. NGC 4278 shows no X-ray halo,
and as could be seen in Figure 10, the diffuse soft X-ray emission
in the central region seems to trace cold and warm gas and is too
weak to serve as an efficient energy reservoir in order to explain
the observed strong optical and infrared emission.

Based on SAURON results, Sarzi et al. (2010) suggest that warm gas
in early-type galaxies are generally powered by evolved stellar
sources, especially post AGB stars, based on their calculation that
an old stellar population is able to provide sufficient ionizing
photons for the observed warm gas, as well as the fact that surface
brightness of H$_{\beta}$ emission is tightly correlated with that
of local optical continuum emission. The latter phenomenon was
previously pointed out by (Macchetto et al. 1996). In Sarzi et al.
(2010), the results have revealed the unusual ionization structure
of NGC 4278. However, the dominant ionization source in NGC 4278
could still be photoionization by UV radiation from old stars
instead of fast shock, based on the fact that high ionization
regions are characterized by low surface brightness of H$_{\beta}$
emissions, which contradicts with common case for shock-heated
region. They also argued that the shock velocity predicted by shock
models ($300-500km/s$, depending on whether to introduce a HII
precursor) is too large to maintain a stable rotational pattern as
seen in NGC 4278.

Nevertheless, dynamical environment giving birth to
hundred-to-thousand pc scale shock is not unexpected for the center
of NGC 4278. As shown by Sarzi et al. (2006), the kinematic of
ionized gas is different from that of stars, with its position angel
of rotation axis showing a maximum deviation from that of the stars
at the outmost region, while this misalignment angle between stars
and gas gradually decreases toward the inner region and reach a
minimum at the galaxy center. This kinematic feature is not
difficult to understand. Morganti et al. (2006) point out that the
central rotational structure of warm gas in NGC 4278 seems to be
just a sub component of the large extended HI disk. It is commonly
recognized that massive cold gas disk in elliptical galaxies is
usually a product of merging with a gas-rich galaxy. The faint
tail-like features in HI map of NGC 4278 also support this idea.
Therefore, shock triggered by random collisions between gas clouds
could possibly happen when gas is accreted from the extended HI disk
to the central stellar gravitational well. It is worthwhile to
notice that, the western high ionization region shown in Figure 3
exactly corresponds to area in velocity map where gas experience
turning of position angle (Sarzi et al. 2006).

Our observational results reveal several possible signs for a shock
process happening in the central region of NGC 4278. As discussed
above, the extended strong emission of [SiII] could explained
by sputtering destruction of silicate dust grain in the fast shock
process instead of X-ray photoionization by central AGN, this is
further supported by possibly modified size distribution of PAHs, as implied
by different PAH band ratios. Recently, Kaneda et al. (2010) show that in NGC 4589, another PAH-detected 
ellipticals sharing similar infrared spectral features with NGC 4278, the PAH17 distribution 
is different from that of PAH11.3 emission. The authors explain this segragation as that 
PAH11.3 arises from newly formed PAHs when dusty gas falling toward the center. 
In Figure 5 we find similar results, which may imply that PAHs are 
undergoing reprocessing in extended area.
 On the other hand, the X-ray diffuse
emission spatially follows the main feature of ionized gas,
molecular and dust. The temperature of thermal component found by
Younes et al. (2010), $0.4 \pm 0.1keV$ within 10'' aperture,
indicates a shock velocity of $\sim 600km/s$ (Draine \& McKee, 1993)
and is higher than expected from optical line diagnostics. However,
the authors also point out that the central diffuse emission may
contain more than one temperature components, with a less-hot
components in the extended region.

We find an excess luminosity of molecular hydrogen rotational lines
populated in high excitation temperature comparing with that found
in PDR regions of starforming galaxies, with respect to aromatic
bands, $24\mu m$ continuum, Total IR luminosity as well as
[SiII]$34.8\mu m$ emission (Roussel et al 2007). This component of extended warm molecular associated with
highly ionized gas also provide a sign for extended shock. Enhanced emissions from
warm molecular is found for AGN objects in SINGs sample and stellar/galactic scale
shock-heated regions. Roussel et al. argue that X-ray excitation
seems unlikely to be responsible for excess warm molecular hydrogen
in SINGs AGN sub-sample, instead they suggest fast-shock triggered
by dynamical perturbation as the heating source. Furthermore,
we have found an extended high-ionization region related with
enhanced H$_{2}$ S(1) emission. These facts support the idea that
X-ray excitation from central AGN could not explain the observed
excess, high excitation temperature H$_{2}$ emission. A second
heating mechanism could be also supported by the possible deviation
of H$_{2}$ S(1) peak from other lines as shown in Figure 6. On the
other hand, there is no evidence of dynamical perturbation directly
related with the central AGN. While being classified as a FR I radio
galaxy, the radio jet of NGC 4278 is confined to milli-arcsec scale
and shows a direction approximately perpendicular to the elongated
main feature of warm gas (Falcke et al. 2000). The HST STIS high
spacial resolution observations also indicate that the twist of
velocity field only happens within the inner 0.5'' region (Walsh et
al. 2008).

Guillard et al. (2009) develop a common scenario to
interpret the dominance of warm H$_{2}$ emissions in dynamically
disturbed regions, in which pre-shock gas exists in both high
density clouds and low-density intercloud gas. Under a higher shock
velocity, the low-density, diffuse gas is heated to plasma with
temperature a few $\times{10^6}K$, while the dusty clouds heated to
lower than $10^6K$ cool efficiently, dissipate their kinematic
turbulent energy into H${_2}$ emission through slow-shock. This
scenario is consistent with our finding that the distribution of
H$_{2}$ emission and PAH emissions are similar, while both of which
depart from ionized emissions. It provides a natural explanation for
the anti-correlation between $8\mu m$ PAH feature and
[OIII]$\lambda5007$/H$_{\beta}$ shown in Figure 3, since PAHs is
expected to be entirely destructed in fast-shock heated plasma
(Micelotta et al. 2010). If so, the decreased [SiII]/[SIII] ratio in
this region might be a result of relatively more pre-shock gas
distributed in high density clouds.

It is uncertain, however, whether the shock component revealed by
above evidences could be responsible for the bulk of gas ionization
in the central elongated feature. The optical and infrared fine
structure line diagnostics allowed by current data are not
sufficient to clearly distinguish between UV photoionization and
fast-shock. Although there are abundant optical spectral
observations for NGC 4278 (Rose et al. 1984; Ho et al. 1997; Walsh
et al. 2008), these observations are usually based on slit centered
at the central AGN. We notice that due to detectable high excitation
line [OIII]$\lambda4363$ in its optical slit spectrum, NGC 4278 were
suggested as an possible example of shock-excited LINERs (Rose et
al. 1984). An obstacle to view shock-heating as a main ionization
mechanism is the low surface brightness of high ionization area.
Interestingly, we find a discrepancy between fine-structure lines
with different ionization potentials, with the highest ionization
mapped line, [NeIII]$15.6\mu m$, showing the most symmetrical
distribution. A much lower [NeIII]/[NeII] is expected to be produced
in shock region than in gas ionized by FUV radiation (Spinoglio \&
Malkan, 1992; Allen et al. 2008) , if this feature is further
confirmed, it implies that shock-heating may play a role in
excitation of lower ionized species in extended region.

\section{Conclusion}

We performed the Spitzer IRS spectral mapping observations toward
the central region of the nearby elliptical galaxy, NGC 4278. We
discovered rich emission features produced by warm dust, molecular
hydrogen and ionized gas in the galaxy center. The multi-phases of
warm and hot gas generally coincide well with optical line
distribution, suggesting they belong to a same structure. We confirm
the reverse distribution of ionization states shown in optical
observations. The spectrum of nuclear region is characterized by
strong [SiII]$34.8\mu m$ emission, which is naturally explained by
reduced silicon depletion through dust sputtering. The PAH band
ratios in the nuclear region could be interpreted by modified size
destruction under selective destruction, especially, the PAH
${7.7\mu m}/{11.3\mu m}$ ratio, decreases while moving away from the
nuclei, indicating a high fraction of neutral PAHs or a deficiency
of small PAHs in the outer region. The warm molecular hydrogen shows
excessive emissions arising from pure rotational transitions with
respect to both dust and ionized emission lines, excitation diagram
of pure rotational lines gives high excitation temperature similar
to some galactic shock-heated regions. We found a high ionization
extended region associated with enhanced rotational molecular
emission. We conclude that a hundred-to-thousand pc scale shock
region triggered by interaction between clouds and diffuse cold gas
accreted from the outer region of the galaxy should exist under the
dynamically perturbed environment in the center of NGC 4278, while
the contribution of this shock component to the total ionization of
gas is still unclear.

\section{Acknowledgements}

The authors are very grateful to Marc Sarzi and Hidehiro Kaneda for
providing us SAURON data of NGC 4278 and IRS spectra of quiescent
elliptical galaxies. This research made use of Tiny Tim/Spiter, developed by John Krist for the
Spitzer Science Center. The Center is managed by the California Institute of Technology under a 
contract with NASA. We would also like to acknowledge the financial
support from the Natural Science Foundation of China under grants
10878010 and 10633040, and  the National Basic Research Program (973
program No. 2007CB815405). This research has made use of NASA's
Astrophysics Data System Bibliographic Services and the NASA/IPAC
Extragalactic Database (NED), which is operated by the Jet
Propulsion Laboratory, California Institute of Technology, under
contract with the National Aeronautics and Space Administration.
This work is based on observations made with the Spitzer Space
Telescope, which is operated by the Jet Propulsion Laboratory,
California Institute of Technology, under NASA contract 1407.


\begin{thebibliography}{99}

\bibitem[]{141} Allen M. G., Groves B. A., Dopita M. A., Sutherland R. S., Kewley L. J.,
2008, ApJS, 178, 20


\bibitem[]{109}
    Binette L., Magris C. G., Stasi��ska G., Bruzual A. G., 1994, A\&A, 292,
    13

\bibitem[]{126}
    Bregman J. N., Temi P., Bregman J. D., 2006, ApJ, 647, 265

\bibitem[]{143} Cluver, M. E., Appleton, P. N., Boulanger, F., et al. 2010, ApJ, 710, 248

\bibitem[]{117} Combes F., Young L. M., Bureau M., 2007, MNRAS, 377, 1795

\bibitem[]{130} Dale D. A. \& Helou G., 2002, ApJ, 576, 159

\bibitem[]{142} Dale D. A., et al., 2009, ApJ, 693, 1821

\bibitem[]{110}
    de Jong T., Norgaard-Nielsen H. U., Jorgensen H. E., Hansen L., A\&A, 232,
    317

\bibitem[]{112} Dopita M. A., Sutherland R. S., 1995, ApJ, 455, 468

\bibitem[]{146}
    Draine B. T. \& McKee C. F., 1993, ARA\&A, 31, 373


\bibitem[]{135}
    Everett M. E. \& Pogge R. W., 1997, IAUS, 182, 106

\bibitem[]{111} Fabian A. C., 1994, ARA\&A, 32, 277

\bibitem[]{147}
    Falcke H., Nagar N. M., Wilson A. S., Ulvestad J. S., 2000, ApJ,
    542, 197

\bibitem[]{124} Fazio G. G. et al., 2004, ApJS, 154, 10

\bibitem[]{108} Filippenko A., 2003, in ASP Conf. Ser. 290, Active Galactic Nuclei: From
Central Engine to Host Galaxy, ed. S. Collin, F. Combes, \& I.
Shlosman (San Francisco: ASP), 369

\bibitem[]{128}
    Galliano F., Madden S., Tielens A. et al., 2008, ApJ, 672, 214

\bibitem[]{128}
        Giveon U., Sternberg A., Lutz D., Feuchtgruber H., Pauldrach A. W. A., 2002, ApJ, 566,
        880

\bibitem[]{114} Goudfrooij P., de Jong T., Hansen L., Norgaard-Nielsen H. U., 1994,
MNRAS, 271, 833

\bibitem[]{127}
    Greggio L. \& Renzini A., 1990, ApJ, 364, 35

\bibitem[]{145} Groves B., Nefs B., Brandl B., 2008, MNRAS, 391, 113

\bibitem[]{148}
    Guillard P., Boulanger F., Pineau D. F. G., Appleton P. N.,
    2010, A\&A, 502, 515

\bibitem[]{108} Ho L. C., Filippenko A. V., Sargent W. L. W., 1997, ApJS, 112, 315

\bibitem[]{136} Herbst, T. M., Graham., J. R., Tsutsui, K., Beckwith, S., Matthews, K., Soifer, B. T. 1990, AJ, 99, 1773

\bibitem[]{139}
    Hewitt J. W., Rho J., Andersen M., Reach W. T., 2009, ApJ, 694, 1266

\bibitem[]{133} Hollenbach D. \& McKee C. F., 1989, ApJ, 342, 306

\bibitem[]{145} Jeong H. et al., 2009, MNRAS, 398, 2028

\bibitem[]{127} Kaneda H., Onaka T., Sakon I., Kitayama T., Matsumoto H., Suzuki S.,
2008, ApJ, 684, 270

\bibitem[]{103} Knapp G., Guhathakurta R., Kim D.-W., Jura M. A., 1989, ApJS, 70, 329

\bibitem[]{113} Knapp G. R., Turner E. L., Cunniffe P. E., 1985, AJ, 90, 454

\bibitem[]{104} Lauer T. R., Faber S. M., Gebhardt K., et al., 2005, AJ, 129, 2138

\bibitem[]{109}
    Macchetto F., Pastoriza M., Caon N., Sparks W. B., Giavalisco M., Bender R., Capaccioli
    M., 1996, A\&AS, 120, 463

\bibitem[]{134}
    Martin P. G., Schwarz D. H., Mandy M. E., 1996, ApJ, 461, 265

\bibitem[]{149}
    Micelotta E. R., Jones A. P., Tielens A. G. G. M., 2010, A\&A,
    510, 36

\bibitem[]{101}Morganti R., de Zeeuw P. T., Oosterloo T. A., McDermid R.
M., Krajnovic D., Cappellari M., Kenn F., Weijmans A., Sarzi M.
2006, MNRAS, 371, 157

\bibitem[]{132}
    Morton D. C. \& Dinerstein H. L., 1976, ApJ, 204, 1

\bibitem[]{138} Neufeld D. A., et al., 2006, ApJ, 649, 816

\bibitem[]{138} Neufeld D. A., Hollenbach D. J., Kaufman M. J., Snell R. L., Melnick G. J., Bergin E. A., Sonnentrucker
P., 2007, ApJ, 664, 890

\bibitem[]{102} Noordermeer Edo, 2006, PhD thesis, Rijksuniversiteit Groningen

\bibitem[]{129} O'Dowd M. J., 2009, ApJ, 705, 885

\bibitem[]{140} Ogle P., Antonucci R., Appleton P. N., Whysong D.,
ApJ, 668, 699

\bibitem[]{115}
    Oosterloo T. A., Morganti R., Sadler E. M., Vergani D., Caldwell
    N., 2002, AJ, 123, 729O

\bibitem[]{149}
    Rose J. A. \& Tripicco M. J., 1984, ApJ, 285, 55

\bibitem[]{106} Sage L. J., Welch G. A., Young L. M., 2007, ApJ, 657, 232

\bibitem[]{101} Sarzi M., 2006, MNRAS, 366, 1151

\bibitem[]{120} Sarzi, M. et al., 2010, MNRAS, 402, 2187

\bibitem[]{107} Serra P., Trager S. C., Oosterloo T. A., Morganti,
R., 2008, A\&A, 483, 57

\bibitem[]{119}
    Smith J. D. T., Armus L., Dale D. A., Roussel H., Sheth K., Buckalew B. A., Jarrett T. H., Helou G., Kennicutt R. C.
    Jr., 2007a, PASP, 119, 1133

\bibitem[]{123} Smith J. D., et al., 2007b, ApJ, 656, 770

\bibitem[]{110}
    Sparks W. B., Macchetto F., Golombek D., 1989, ApJ, 345, 153

\bibitem[]{149} Spinoglio L. \& Malkan M. A., 1992, ApJ, 399, 504

\bibitem[]{137} Roussel H., et al., 2007, ApJ, 669, 959

\bibitem[]{121}
    Tang Y. -P., Gu, Q.-S., Huang, J.-S., Wang, Y.-P., MNRAS, 397,
    1966

\bibitem[]{118} Temi P., Brighenti F., MathewsW. G., Bregman J. D., 2004, ApJS, 151, 237

\bibitem[]{131}
    Temi P., Brighenti F., Mathews W. G., 2009, ApJ, 695, 1

\bibitem[]{121} Terashima Y., Wilson A. S., 2003, ApJ, 583, 145

\bibitem[]{116}
    Tonry J. L., Dressler A., Blakeslee J. P., Ajhar E. A., Fletcher A. B., Luppino G. A., Metzger M. R., Moore C.
    B., 2001, ApJ, 546, 681

\bibitem[]{147}
    Walsh J. L., Barth A. J., Ho L. C., Filippenko A. V., Rix H. -W., Shields J. C., Sarzi Marc., Sargent W. L.
    W., 2008, AJ, 136, 167

\bibitem[]{106} Wiklind T., Combes F., Henkel C., 1995, A\&A, 297, 643

\bibitem[]{127}
    Wu H., Cao C., Hao C. -N., Liu F. -S., Wang J. -L., Xia X. -Y., Deng Z. -G., Young C.
    Ke-Shih, 2005, ApJ, 632, 79

\bibitem[]{122}
    Younes G., Porquet D., Sabra B., Grosso N., Reeves J. N., Allen M.
    G., 2010, arXiv1004.5134

\end{thebibliography}
\end{document}